# A Scintillator Attenuation Spectrometer For Intense Gamma-Rays

E. Liang[1,a], K. Q. Zheng[1], K. Yao[2], W. Lo[1], H. Hasson[3], A. Zhang[2], M. Burns[1], W.H.Wong[4], Y.Zhang[4], A. Dashko[5], H. Quevedo[5], T. Ditmire[5], G. Dyer[6]

[1]Rice University, Physics and Astronomy Department, Houston, TX 77005 USA
[2]California Institute of Technology, Division of Physics, Mathematics and Astronomy, Pasadena, CA 91125 USA
[3]University of Rochester, Physics Department, Rochester, NY 14627 USA
[4]M.D. Anderson Cancer Center, Diagnostic Imaging Division, Houston, TX 77005 USA
[5]University of Texas at Austin, High Energy Density Science Center, Austin, TX 78712 USA
[6]SLAC National Accelerator Laboratory, Linac Coherent Light Source, Menlo Park, CA 94025 USA

[a] Author to whom correspondence should be addressed: liang@rice.edu

## ABSTRACT

A new type of compact high-resolution high-sensitivity gamma-ray spectrometer for short-pulse intense gamma-rays (250 keV – 50 MeV) has been developed by combining the principles of scintillators and attenuation spectrometers. The first prototype of this scintillator attenuation spectrometer (SAS) was tested successfully in Trident laser experiments at LANL. Later versions have been used extensively in the Texas Petawatt laser experiments in Austin TX, and more recently in OMEGA-EP laser experiments at LLE, Rochester, NY. The SAS is particularly useful for high-repetition-rate laser applications. Here we give a concise description of the design principles, capabilities and sample preliminary results of the SAS.

## I. INTRODUCTION

Conventional gamma-ray spectrometers use the "single photon counting" method by employing scintillators (NaI, CsI, BGO)[1][2][3] coupled to photo-multiplier tubes (PMT)[4], or solid-state (e.g. HPGe) detectors [5]. Each gamma-ray deposits all of its energy in the scintillator or solid-state detector, converting its energy into optical photons and/or photoelectrons, which are then amplified for electronic current readout. The gamma-ray spectrum is built up one photon at a time by measuring the total energy deposited by each gamma-ray [6]. In this approach consecutive gamma-rays arriving at the detector must be separated in time longer than the scintillation time, typically > nanoseconds. Otherwise multiple low-energy gamma-rays cannot be distinguished from a single high-energy gamma-ray. Hence single-photon-counting gamma-ray spectrometers cannot be used in short-pulse intense laser experiments, where a large number of gamma-rays arrive in ps to fs time scales, much shorter than the scintillation or electronic readout time.

At present, filter-stack attenuation spectrometer (FSS), made up of a series of high-Z filters arranged in tandem [7] or configured as step wedges, combined with radiographic films or image plates, has been the main diagnostic used to measure continuum gamma-ray spectrum in high-energy-density (HED) and laser experiments. However, FSS allows only a small number of energy channels (typically < 20), and covers energies only up to ~ 6 MeV. For gamma-rays with energy ≥ ~5 MeV, the mass attenuation coefficient reverses its decline due to the photoelectric effect and rises with increasing energy due to pair production [8]. Hence above ~ 5 MeV the attenuation length can no longer provide an unambiguous measure of the gamma-ray energy, even though detailed Monte Carlo simulations based on iterative forward folding methods can provide some constraints on analytic spectral models up to ~ 10 MeV. Other techniques that have been attempted to measure the energy of short-pulse intense gamma-rays include nuclear activation thresholds [9] and Forward Compton scattering (FCS) [10] [11]. However, none of these techniques have been able to provide high-resolution high-sensitivity spectroscopy of intense gamma-rays emitted in short-pulse laser and HED experiments.

Over the past decade, a collaboration between Rice University and the medical imaging group at the MD Anderson Cancer Center (MDACC) in Houston, has developed a new type of gamma-ray spectrometer, which we will call scintillation attenuation spectrometer (SAS). The basic idea is to image the 2-dimensional (2D) scintillation light pattern emitted by a finely pixelated (with ~mm-sized pixels) scintillator matrix when it is irradiated sideways by a narrowly collimated beam of gamma-rays. Since the 2D energy deposition pattern in the scintillator block varies with incident gamma-ray energy, the high-resolution 2D scintillation light patterns can in principle be used to reconstruct the incident gamma-ray spectrum, provided the emerging scintillation light profile is sufficiently bright and faithfully reproduce the local energy deposition profile of the gamma-rays. For this technique to work well, a large number of conditions must be met. Some of the most important requirements are listed here. (a) Each pixel must internally reflect 100% of scintillation photons emitted in that pixel except at the front surface, so that the light emerging from the front surface of that pixel faithfully represents the energy deposited only in that pixel. (b) Each pixel must be as narrow as possible in order to maximize the spatial resolution and minimize the internal absorption of the scintillation light. (c) The scintillation material should have the highest light output for each MeV of absorbed gamma-ray energy. (d) Scintillator material of the highest-Z and highest-density should be used so that the gamma-ray is attenuated efficiently, in order to make the detector as compact as possible. Guided by these four principles and the optimization of many other conflicting requirements, we created a working prototype of the SAS after many trials and errors, using 1.5mm x 1.5mm x 10mm Ce-doped LYSO [12] scintillator crystals for each pixel. Each pixel was fully coated with 3M ESR 0.08mm-thick polymer reflectors on 5 sides except the front surface. We used DYMAX UV-curing adhesive for glue. The cutting, polishing, coating and gluing processes were highly labor intensive and required many specialized equipment developed at MDACC. Ce-doped LYSO was the scintillator of choice because of its high-Z, high-density (7.2 gm/cc) and high light output (~33000 420nm blue photons per MeV absorbed energy) [10] and it is non-hydroscopic. The first SAS prototype, consisting of a 24 x 36 matrix of 1.5mm x1.5mm x 10mm pixelated LYSO(Ce) crystals was successfully demonstrated in the summer of 2015 in a high-intensity Trident short-pulse laser experiment at LANL. Despite the low quality of the first SAS images obtained in the Trident experiments due to the crude CCD camera and light leakage from the container box, this proof-of-principle experiment demonstrated the utility and functionality of the SAS in short-pulse laser experiments. This ground breaking experiment at Trident demonstrated several important properties which provided confidence in the successful construction of a high-resolution compact spectrometer for laser, fusion and many other HED gamma-ray applications. (a) The LYSO crystals produced abundant light that can be easily imaged using conventional CCD cameras without intensification or cryogenics. (b) The gamma-rays penetrated all 36 longitudinal LYSO pixels. This means that a larger crystal matrix block incorporating more pixels can still produce sufficient light output from many more pixels. (c) The 2D light patterns produced by different incident gamma ray energies were clearly distinguishable.

(d) The intense EMP and neutron flux from the Trident experiments did not affect the SAS performance.

We note that pixelated scintillators coupled to PMTs have been widely used for high-resolution gamma-ray imaging in many fields, such as PET cameras in medical imaging, and in high-energy physics experiments [13]. However, their use for spectroscopic measurements of short-pulse intense gamma-rays has not been systematically exploited. Recently, Belm et al [14] explored the use of large segmented blocks of NaI and CsI scintillators to measure gamma-ray spectrum in LWFA laser experiments. However, these detectors were designed for gamma-rays with higher energies (~100 MeV-GeV range). Because large scintillator blocks of NaI and CsI (≥ cm) were used, they resulted in a large (~meter-sized) detector, low light intensity and low spectral resolution. Typically, only a simple model spectrum with a few input parameters (e.g. an exponential or power law function) can be meaningfully constrained using iterative Monte Carlo simulations, by matching the observed light profiles with GEANT4 [15] model predictions [14]. To our knowledge no previous finely-pixelated scintillator matrix spectrometer has been designed or constructed to provide the model-independent reconstruction of an unknown incident gamma-ray spectrum, with high spectral resolution and high signal-to-noise. This is the goal of the SAS.

## II. DESIGN, DATA SAMPLE AND CALIBRATION

After the successful proof-of-principle demonstration with the 2015 Trident laser experiments, we constructed 36x48 (Fig.1) and 36x60 LYSO matrix blocks and used them in our Texas Petawatt (TPW) laser experiments in 2016 and 2018. The SAS was also used successfully in OMEGA-EP experiments in 2020. A 36x60 LYSO matrix block, measuring only 6 cm x 8 cm, can fully capture the scintillation light pattern of gamma-rays up to 50 MeV. We have upgraded the CCD camera to a high-sensitivity high-speed non-cryogenic camera with wide-field non-distorting lens and a CCD chip optimized for 420 nm light, plus remote control and data-link. The entire SAS apparatus is housed in a light-tight black box approximately the size of a shoe box (Fig.2), with thick (> 2 inches) lead shielding all around and a 3 mm - 6 mm pinhole for the collimation of incident gamma-rays, which irradiate the matrix block along the central long axis. A tiny hole in the SAS housing is used for alignment of the matrix block central long axis with the laser target. In some high neutron flux experiments, additional neutron-shielding material is used to protect the front end of the housing around the LYSO block. After using the SAS in many TPW shots, we have found that the CCD camera itself is insensitive to EMPs. However, the cable connecting the CCD camera to the control computer (located far from the target chamber) must still be protected with ferrite coils to avoid upsetting the computer. Our current CCD camera is capable of taking repeated 1ms exposures. Hence the SAS is ideal for use in high-repetition-rate laser experiments.

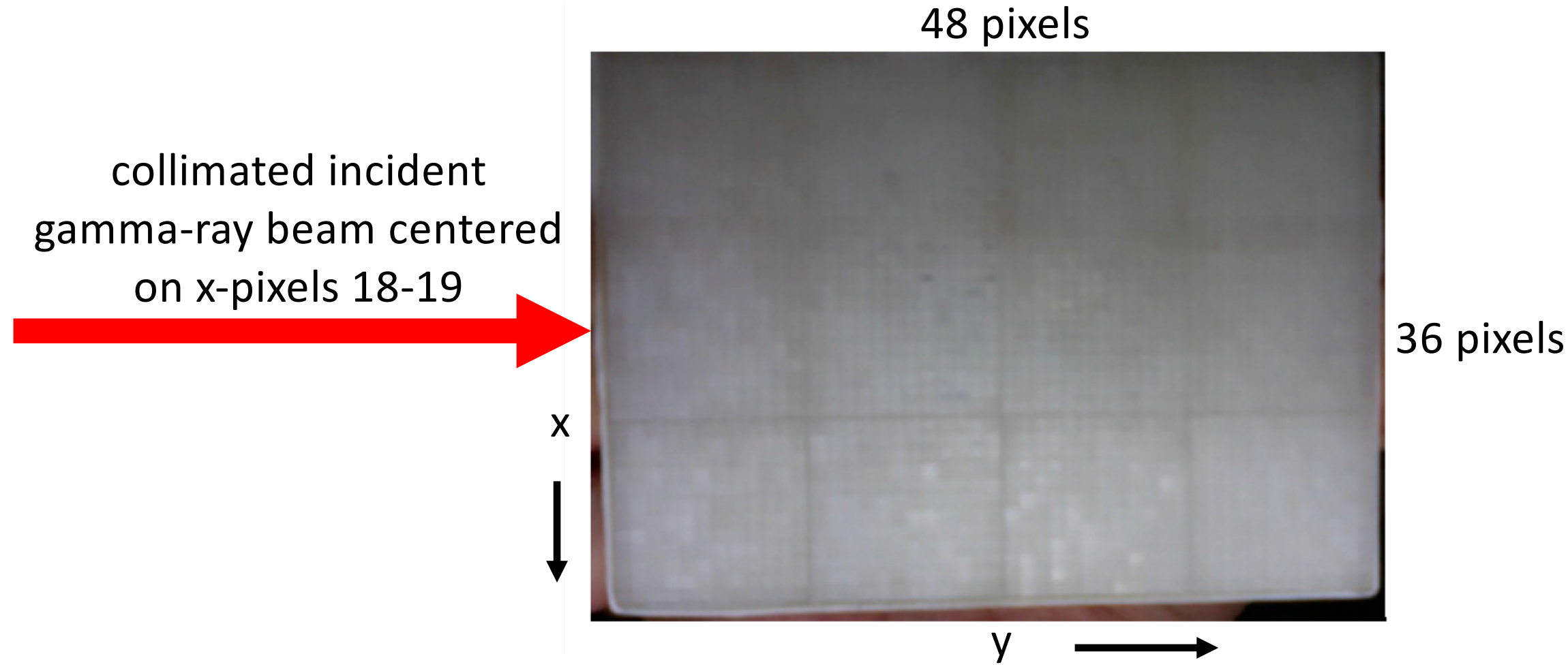

**FIG.1** Picture of a pixelated LYSO crystal block used in the SAS, consisting of 36 x 48 1.5mm x1.5mm x10mm pixels. The entire block which measures 6 cm in height (labelled x) x 8 cm in length (labeled y) x 1 cm in depth (into the picture) is made of 3 x 4 sub-blocks each of which consists of 12 x 12 pixels. Collimated gamma-rays enter the crystal block from the left at x-pixels 18-19. Following the convention used in later diagrams, we measure the x-coordinate from top to bottom and y coordinate from left to right.

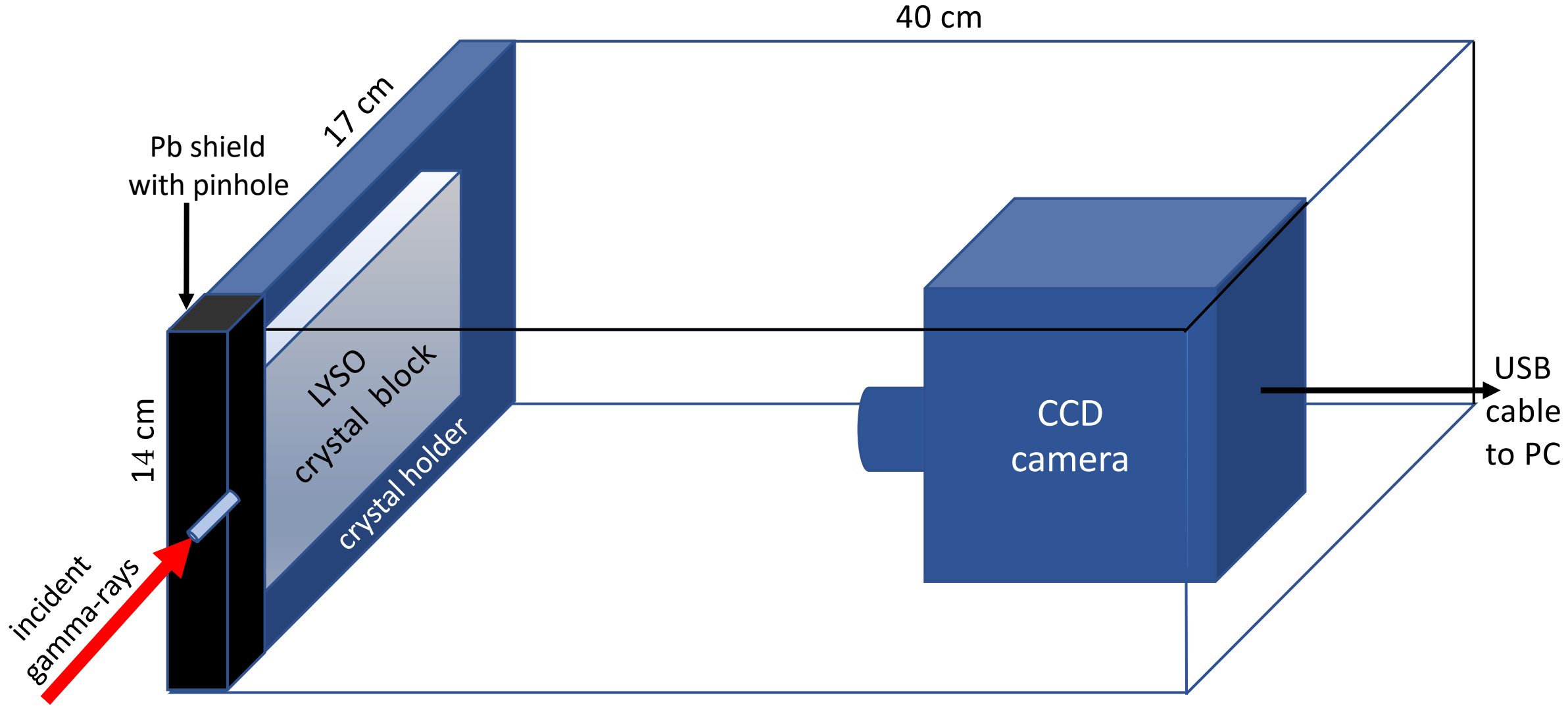

**FIG.2** Artist sketch of the SAS layout with overall dimensions of the light-tight housing. Not shown are thick external lead shields with a 6 mm-diameter pinhole used to collimate the incident gamma-ray beam and protect the crystal from scatter gamma-rays. The lens of the CCD camera is positioned at 17-18 cm from the scintillator front surface to provide the best image of the entire crystal block.

Fig.3 shows a typical SAS raw image taken during the 2016 TPW experimental campaign, using 6mm pinhole collimators. Fig.4 shows raw SAS images at two different detector angles from a single 2018 TPW shot using 3mm pinhole collimators. It demonstrates that the gamma-rays emitted at the target normal (TN) direction is much brighter and harder than those emitted at 90° from target normal.

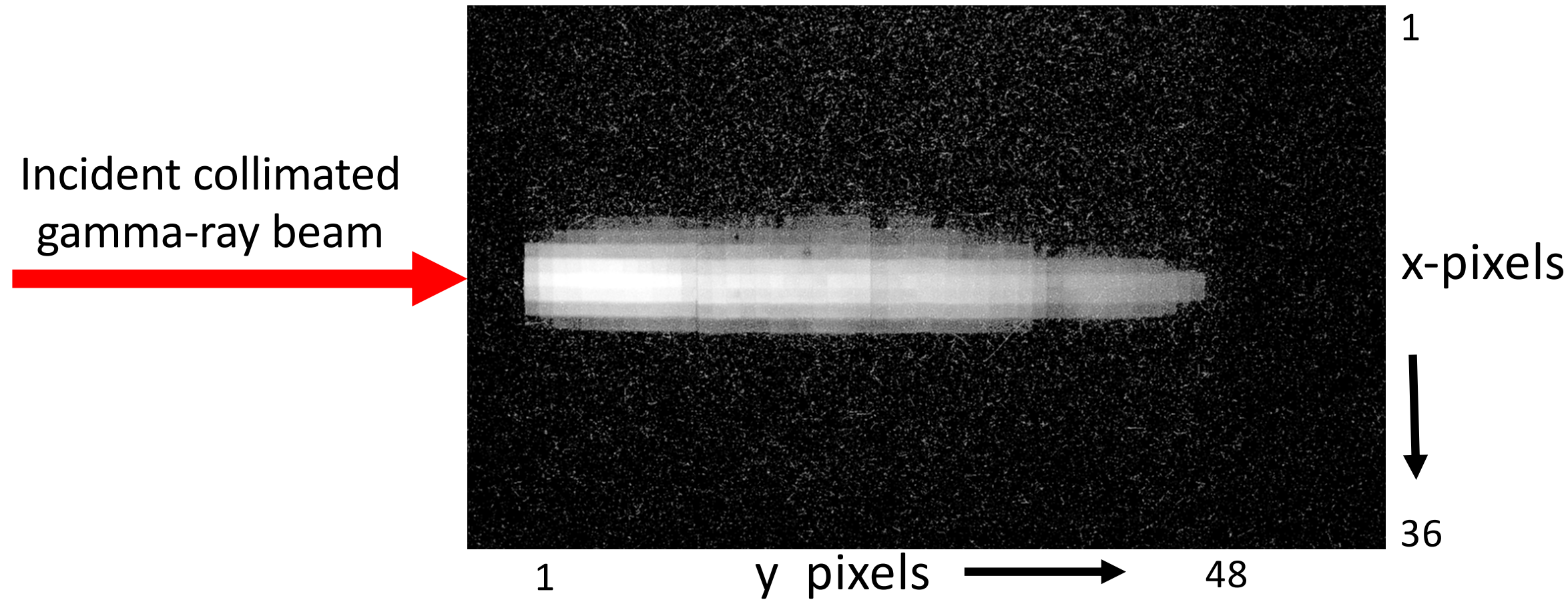

**FIG.3** (a) Raw image of Shot 10084 LYSO scintillation light pattern from 2016 TPW experiments. Gamma-rays entered the crystal block from a 6 mm-diameter pinhole at the center of the left edge. This image shows over 350 visible bright pixels, with all pixels along the central axis luminous except the last pixel. The dark zone to the left of y=1 is part of the internal lead shield and to the right of y=48 is the crystal holder (see Fig.2).

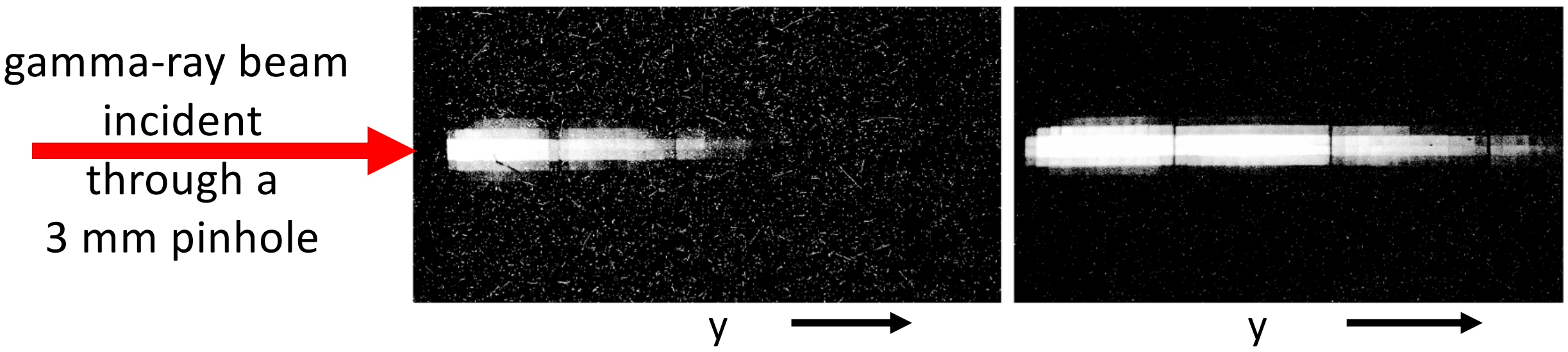

**FIG.4** Two SAS raw images from a 2018 TPW shot using 3 mm-diameter pinholes. The left image comes from a SAS located at ~ 90° from target normal. The right image comes from another SAS located at target normal. These two images clearly demonstrate that the gamma-rays emitted at target normal are brighter and harder than those emitted at ~ 90° from target normal. The CCD camera used to obtain the left picture was set to be 50% more sensitive than the one used to obtain the right picture.

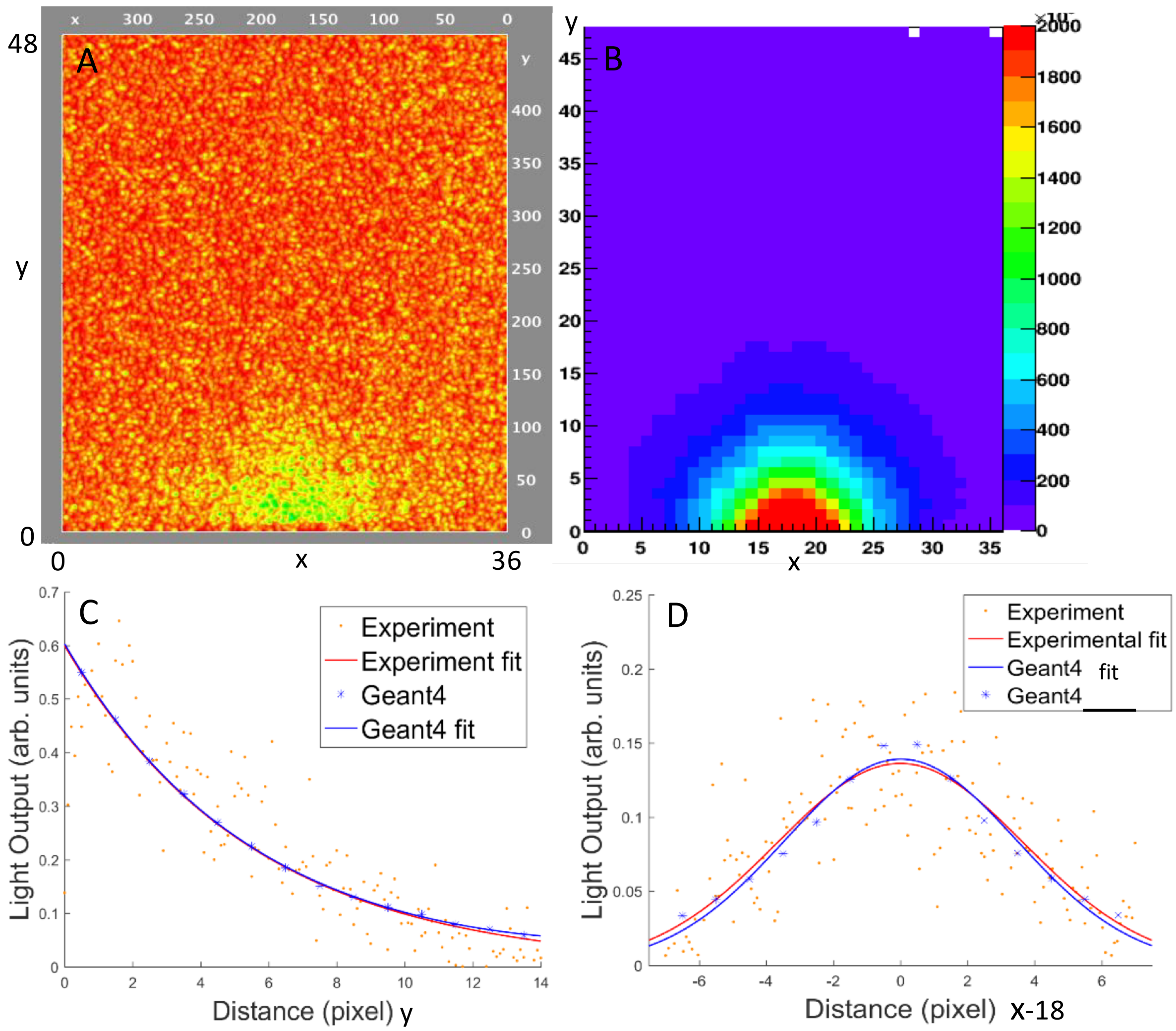

**FIG.5** Calibration of the SAS using a $^{137}$Cs (662 keV) source of known intensity. (a) is raw SAS image, (b) is GEANT4 simulated image, (c) is the SAS longitudinal light profile along y-axis compared to the corresponding GEANT4 simulated profile, (d) is the SAS transverse light profile along x-axis compared to the corresponding GEANT4 simulated profile. The experimental and GEANT4 curves are peak-aligned. The agreement between experimental data and GEANT4 predictions are excellent.

We have calibrated the SAS response in the laboratory using gamma-ray line-emitting isotopes such as $^{137}$Cs, $^{22}$Na and $^{207}$Bi. $^{137}$Cs was our strongest and cleanest point source. Even though our $^{22}$Na and $^{207}$Bi calibration results basically agree with GEANT4 predictions, their signals are much weaker, and the sources are more complicated and difficult to model with GEANT4 input. Hence we only show the results of $^{137}$Cs here. Fig.5 shows the SAS image for $^{137}$Cs (0.662 MeV) compared to the GEANT4 [15] simulated image (see Sec.3 below) using a detailed model of the LYSO crystal block, including the lead shielding and pinhole. We see that the SAS data agree very well with GEANT4 predictions. This validates the GEANT4 simulation model and results at least for low-energy gamma-rays. In Sec.6 we discuss future plans for additional calibration experiments using higher energy sources.

## III. GEANT4 MODELING OF LYSO LIGHT PATTERNS

We use GEANT4 [15] to model the response of our LYSO matrix block to incident gamma-rays. GEANT4 is the industry-standard particle physics code which incorporates almost all relevant physics of gamma-ray interaction with matter. GEANT4 employs the Monte Carlo method, which creates and tracks each particle (photons, electrons, positrons) after emission and scattering events, and eliminates the particle after an absorption event. Interactions are based on their energy-angle-dependent mean-free-paths. Below we summarize the model inputs.

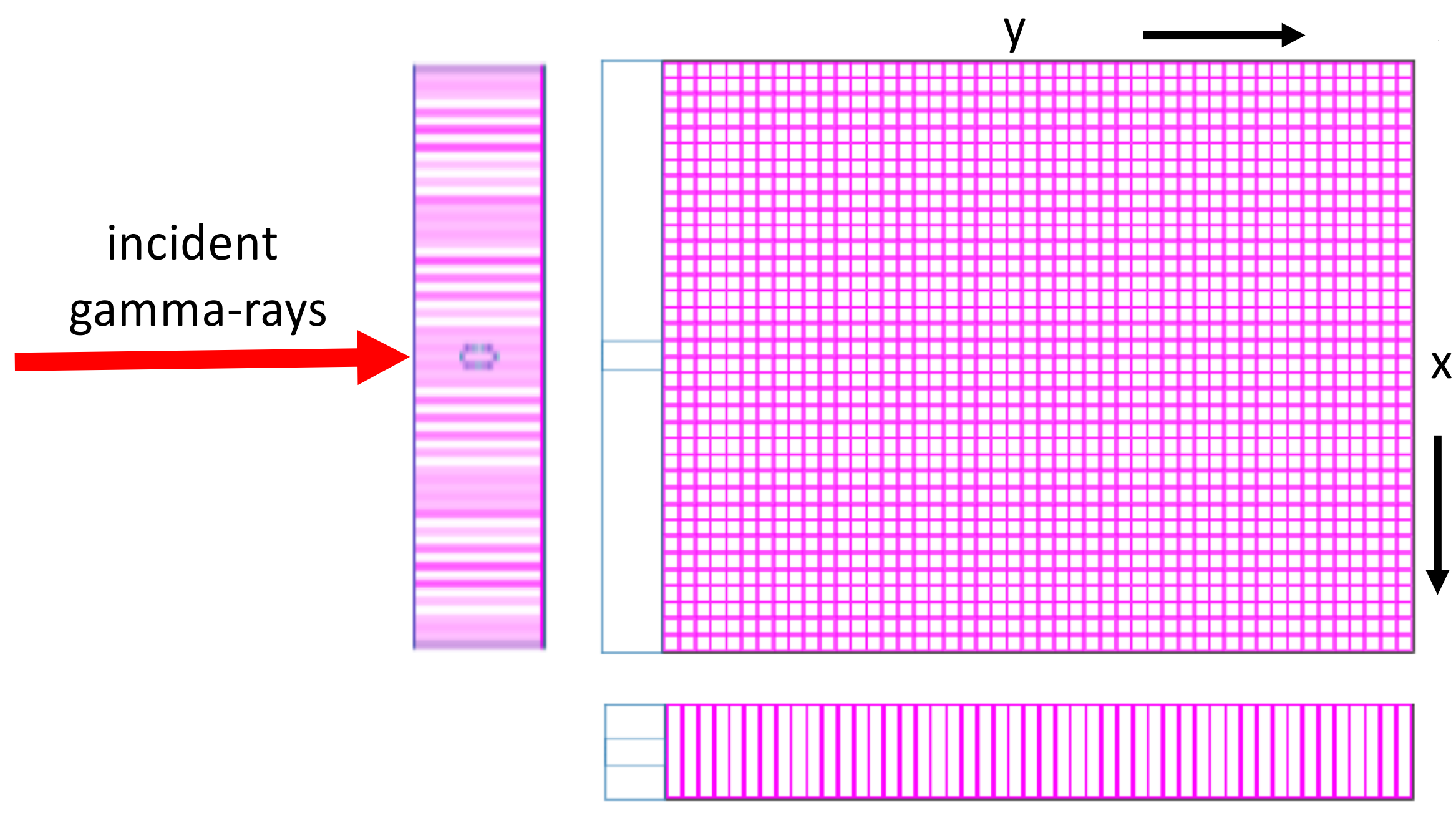

**FIG.6** Geometry of the LYSO matrix plus lead shields used in our GEANT4 simulations. The bottom panel depicts the depth of each pixel. The lead collimator on the left is only included in the GEANT4 input for the $^{137}$Cs calibration experiment of Fig.5. To model laser experiments a cylindrical beam of gamma-rays is injected directly into the matrix along the central axis.

### A. Geometry and Material Input

In GEANT4, the entire scintillator block is modeled as an array of 36 x 1 x 48 LYSO pixels individually coated with 3M ESR reflector of thickness =0.008 cm. Each LYSO pixel including reflector and glue has a combined dimension of 0.16667cm x 1.0cm x 0.16667cm. The scintillator medium is $Lu_{1.9}Y_{0.1}SiO_5$ doped with 0.5% Ce with a density of 7.4 g/cm$^3$. The 3M ESR reflector is made of $C_5H_8O_2$ with a density of 1.18 g/cm$^3$. The chemical composition and density of DYMAX UV-curing adhesive was taken from the company website. All known elements and their compositions are included in the GEANT4 model input.

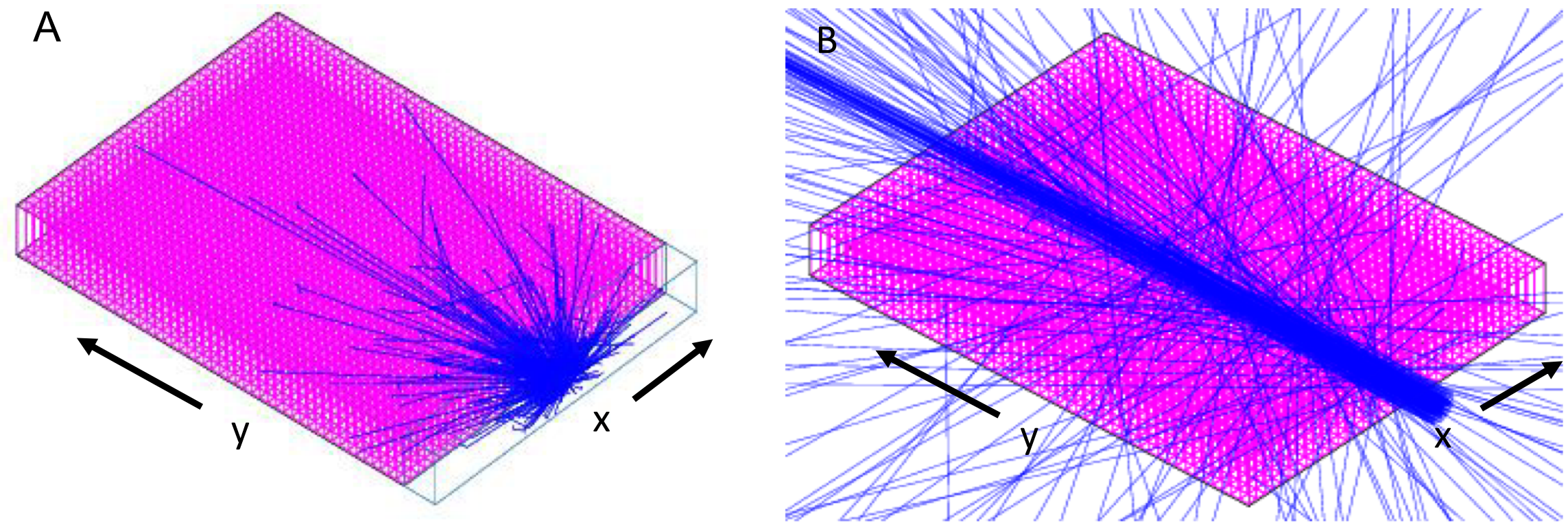

**FIG.7** (a) Sample GEANT4 gamma-ray tracks for the $^{137}$Cs (0.662 MeV) calibration experiment. (b) Sample GEANT4 gamma-ray tracks for a 6mm-diameter cylindrical beam incident along the central axis of the matrix block from the lower right edge (y=0).

**B. Gamma Ray Source Input**

For the calibration experiment of Fig.5, a hemispheric $^{137}$Cs source of 5 million particles at 0.662 MeV is injected and filtered by a lead collimator of 6cm x 0.6cm x 1 cm with a 3mm pinhole (Fig.7a). To model the collimated gamma-ray source from laser experiments, a 6 mm-diameter cylindrical beam of 1 million monoenergetic gamma-rays is injected along the central axis (Fig.7b). The energy deposition in each pixel is computed instead of the actual scintillation light output, because in GEANT4 scintillation calculations, the number of optical photons emitted is directly proportional to local energy deposition. An output file containing the 2D histogram of energy deposition values in all (36 x 48=1728) pixels is generated for each incident gamma-ray energy. Sample gamma-ray tracks from both GEANT4 simulations are illustrated in Fig.7.

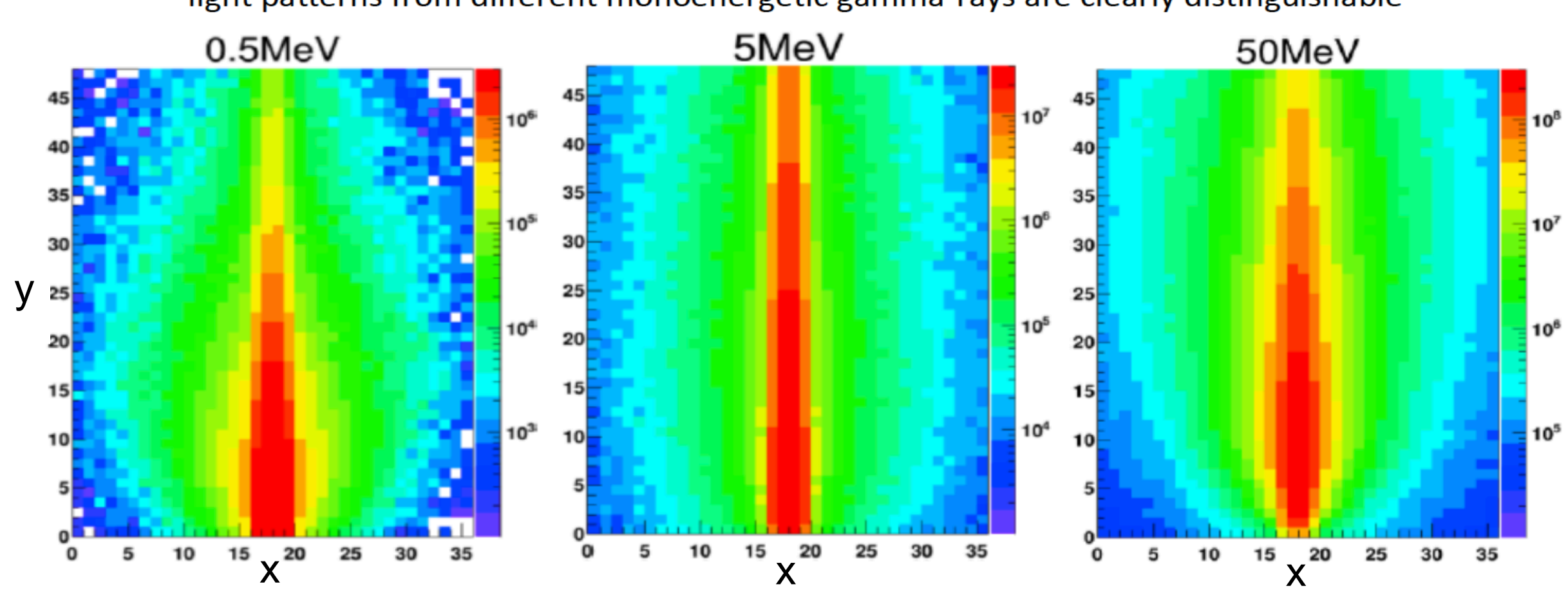

**FIG.8** GEANT4-simulated scintillation light patterns of 36x48 LYSO matrix for three different monoenergetic gamma-ray energies. The 6mm-diameter gamma-ray beam enter the central axis of the matrix from the bottom in all three cases. The color is in log scale.

Fig.8 shows sample GEANT4-simulated LYSO scintillation light patterns of three monoenergetic incident gamma-rays to highlight the change of the light pattern with gamma-ray energy. The "candle light" pattern of 0.5 MeV gamma-rays (left panel) is dominated by the photoelectric effect. The "tear drop" pattern of 50 MeV gamma-rays (right panel) is dominated by pair production. Compton electrons dominate the scintillation light of 5 MeV gamma-rays (middle panel). In this case the light profile is more concentrated along the central axis and declines more slowly with y than in the other two cases (see also Fig.9(c)). These examples highlight the importance of using the full 2D patterns of scintillation light to distinguish different gamma-ray energies.

## IV. CONSTRUCTION OF THE DETECTOR RESPONSE MATRIX

For all gamma-ray spectrometers, the first step to invert or deconvolve the incident gamma-ray spectrum from the detector signal is to build up a detector response matrix (DRM) using Monte Carlo simulations described in Sec.3, which maps the incident gamma-ray energies to the light output of all pixels. For our 36 x 48 matrix, only gamma-rays up to 50 MeV are effectively captured. Hence in our GEANT4 simulations, we inject monoenergetic gamma-rays from 0.25 MeV to 50 MeV at 0.25 MeV intervals, using $10^6$ Monte Carlo particles per run. This allows us to construct a 200x200 DRM, which maps 200 gamma-ray energy channels onto the scintillation light output of 200 "effective" LYSO pixels (Fig.9). As Fig.8 shows, most scintillation light output is concentrated in the five longitudinal columns adjacent to the central axis of the crystal matrix. Hence we order the pixel numbers sequentially as 1-48 for the first column, 49-96 for the second column, 97-144 for the third column, 145-192 for the fourth column. For all of our laser experiments, the light output of the fifth column is relatively faint, hence we combine the 48 pixels of the fifth column into 8 "large" pixels by averaging the light of every 6 pixels. We also average the light output from the corresponding pixels of the left half and right half of the matrix to get better statistics. This gives us the light output of 200 "effective pixels". We note that a 200 x 200 DRM is optimal only for gamma-rays up to 50 MeV. If we want to measure gamma-rays > 50 MeV, we will need more longitudinal pixels and more energy channels, resulting in a larger DRM. Moreover, a square DRM is required only if we want to construct the inverse DRM. The inverse matrix is useful only if the DRM is highly non-singular so that the incident gamma-ray spectrum can be obtained directly by multiplying the light output column vector with the inverse DRM. As we discuss in Section 5 below, the SAS DRM is "almost singular", meaning that any tiny noise in the light output data will be amplified by the inverse matrix to dominate the true signal and give unphysical results. In practice, a rectangular matrix, with more pixels than energy channels (i.e. an over-determined system of linear equations), actually leads to more stable and reliable inversion solutions. These are discussed in Sec.5.

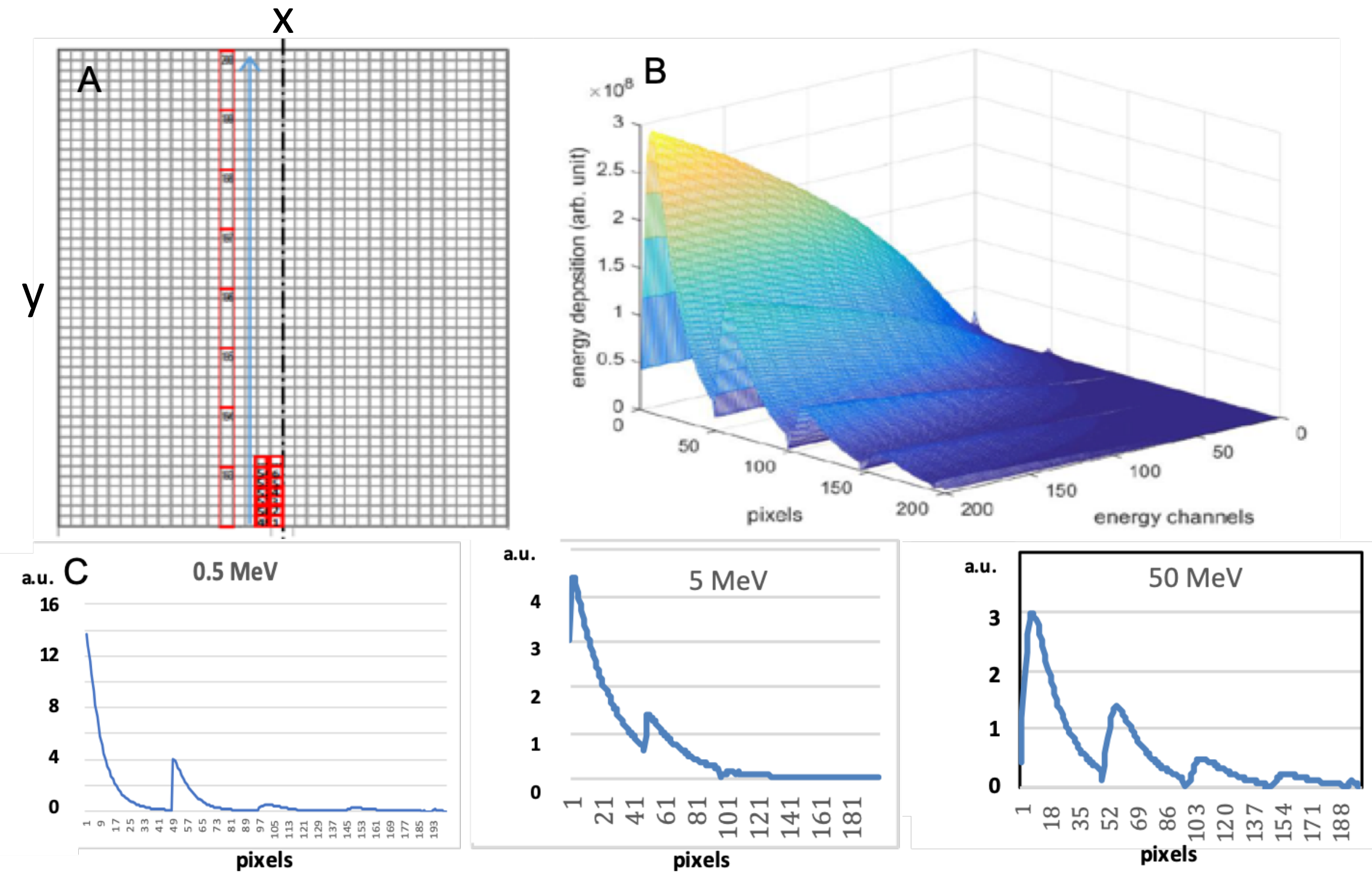

**FIG.9** (a) Sequential labeling of the pixel number starting from the center line of the 36 x 48 LYSO matrix: pixels 1 – 48 denote first column, pixels 49-96 denote second column, etc. In the fifth column we combine six LYSO pixels into 1 large pixel due to the faintness of the light. Gamma-rays enter the matrix from the bottom center (x=18-19). Left and right halves of the matrix are are averaged to produce the DRM. (b) 3D contour plot of the 200 x 200 DRM obtained with 200 GEANT4 simulations. The folded accordion pattern of the DRM is due to the way we order the pixel numbers in Fig.(a). (c) Numerical light profiles corresponding to the three images of Fig.8. y-axis units are arbitrary.

Since the bright pixels are concentrated in the 5 columns closest to the central axis of the matrix (Fig.8), one may wonder why we need 36 transverse pixels given the high cost of LYSO crystals. However, the LYSO crystals produced at MDACC only come in 12x12 sub-blocks (Fig.1) due to the manufacturing process. If we use only a single column of 12 x 12 sub-blocks to form a 12 x 48 matrix, there are not sufficient "dark pixels" outside the bright region to construct a meaningful background for background subtraction. For many practical reasons we have to go far away from both the bright pixel region and the outer edge of the crystal block to obtain a good reliable background. Hence 36 transverse pixels are often needed for background subtraction. Furthermore, if we later want to extend the gamma-ray energies to > 50 MeV, the transverse light profile will spread beyond 10 or even 12 pixels. In that case we will need the additional transverse pixels anyway.

## V. DECONVOLUTION OF GAMMA-RAY SPECTRUM FROM SAS IMAGE

Because the 200x200 SAS DRM matrix is almost-singular with a condition number > $10^7$, direct inversion using the inverse DRM matrix is highly unstable and impractical, since small noise in the light output will be greatly amplified by the inverse matrix to completely dominate the true signal. Over the past few years, we have explored a large variety of regularization and forward folding methods. At present, we have settled on a hybrid algorithm based on a combination of the Non-Negative Truncated Singular Value Decomposition method (TSVDNN [16] with 5 – 6 terms) and the Parametrized Forwarded Folding (PFF) [17] method, which seem to give the best results. When this hybrid algorithm is applied to model input spectrum with both broad spectral features and narrow spectral lines, it gives reliable and stable inversion solutions. When this algorithm is applied to actual SAS data from TPW experiments, it also appears to produce robust stable results. We have also determined that in most cases, the best inversion solutions are obtained by consolidating the 200 energy channels into 100 energy channels, so that the reduced DRM becomes a 200 x 100 rectangular matrix. This reduced 200 x 100 DRM still provides us with 0.5 MeV spectral resolution for the inverted spectrum up to 50 MeV.

Here we provide a concise discussion of the hybrid inversion algorithm. The TSVDNN is based on the truncated singular value decomposition (TSVD) method [17], with a constraint added to the solution space to accept only positive solution values [14]. Singular value decomposition (SVD) [17][18][19] decomposes the solution to the inverse problem into a series of vectors with associated coefficients related to the singular values of the matrix. The SVD vectors are ordered in such a way that the first few terms with the largest singular values are least susceptible to data noise [19]. As a result, the truncation of higher order terms with the smallest singular values, which are most susceptible to noise, leads to a stable solution [19]. In our case we find that keeping only the first 5 – 6 terms of the SVD series gives the best results. However, we find that such brute force truncation of the SVD series frequently leads to inversion solutions with negative values which are unphysical. Through experiments with simulated data, we learn that the negative values are likely cancelled by the truncated terms in the case of perfect data without noise. But the inversion solution becomes dominated by noise if we add back those terms for real data with noise. To circumvent this difficulty, we apply an iteration criterion in which we zero out all the negative values in the TSVD series during each iteration, when we solve for the coefficients of the decomposition vectors. It turns out that this TSVDNN algorithm gives rather good results in reproducing the position of the peaks and the overall shape of the input spectrum. Only when we have a very narrow peak in the input spectrum, then the TSVDNN solution starts to deviate from the true shape of the peak and returns a wider peak. Nevertheless, the position of the original peak is still well captured. We can then resort to our physical knowledge about the truth width of the peaks (e.g. narrow nuclear lines). We emphasize that the TSVDNN solution is completely model independent and requires no prior knowledge of the shape or form of the input spectrum, and is tolerant of data noise.

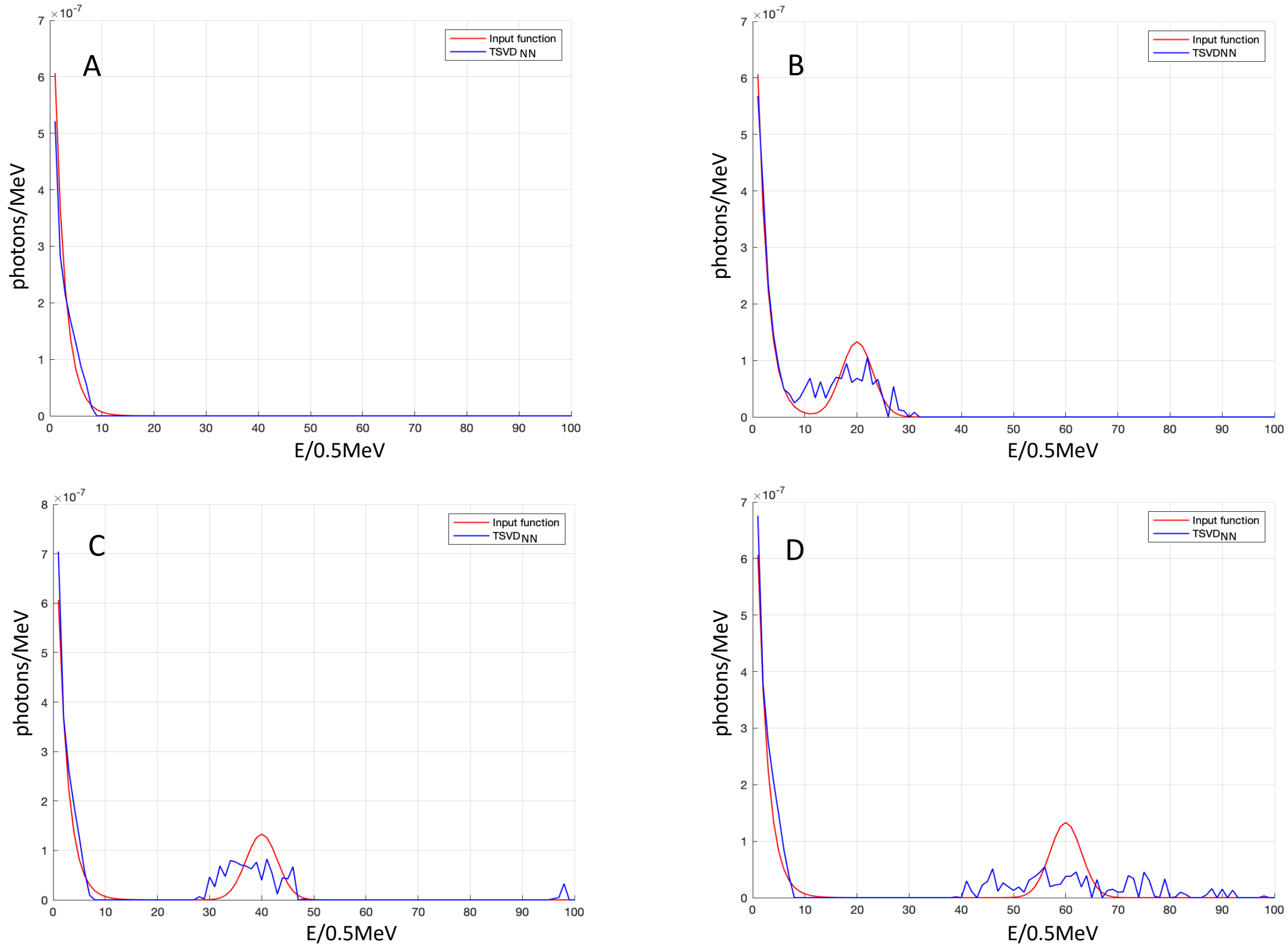

**FIG.10** Examples of TSVDNN inversion solutions to analytic model input spectra. Red curves are the input gamma-ray spectra consisting of a simple exponential function (a), or an exponential function plus a Gaussian bump of various peak energies (b)(c)(d). 5% random noise is added to the model-generated output data before the data is inverted using the TSVDNN algorithm. The TSVDNN solutions are shown as the blue curves.

Fig.10 shows four examples of the TSVDNN solution (blue) for input model spectra (red) consisting of an exponential function (Fig.10(a)), or an exponential function plus a Gaussian bump (Figs.10(b)(c)(d)). The input model spectrum is first convolved with the DRM to form a 200-pixel light output profile. 5% random noise is added to the data before the data is inverted using the TSVDNN algorithm. In all cases the TSVDNN inversion solution reproduces the exponential shape very well. The high-energy bump of the inversion solution gets broader and noisier as its peak energy is increased. But the centroid of the broadened bump remains close to the correct input energy. The area under each spectral component is also well preserved by TSVDNN.

The next step is to improve on the TSVDNN solution with what we call a Parametrized Forward Folding (PFF) [18] technique. We initiate an iterative forward folding procedure with an analytic trial function which is motivated by, and gives the best-fit to, the model-independent TSVDNN solution using $R^2$-minimization. The final PFF parameters are then determined by least-square fitting to the original light profile data using the Levenberg-Marquardt algorithm. In other words, the final PFF solution is not enslaved by the TSVDNN solution, even though the starting values of the PFF iteration was based on the TSVDNN values.

Fig.11 illustrates the application of the hybrid algorithm to an input model spectrum consisting of an exponential function plus a narrow line at 4.5 MeV and a broad Gaussian bump centered at 20 MeV (Fig.11(a)). 5% random noise is added to the simulated light output data (Fig.11(b) black curve) before it is inverted using TSVDNN. Even though this TSVDNN inversion solution is noisy and not a good match to the input spectrum, it still captures the peak energies of both the narrow peak and broad peak correctly. We then follow up with iterative PFF optimization, using an analytic model function based on the TSVDNN solution and starting values obtained from fitting the TSVDNN solution. The final PFF solution (Fig.11(d)) is in good agreement with the original input spectrum. We have performed many other similar exercises using different input model spectra, all with satisfactory results. These exercises give us confidence in applying this hybrid algorithm to deconvolve gamma-ray spectrum from experimental data obtained by the SAS.

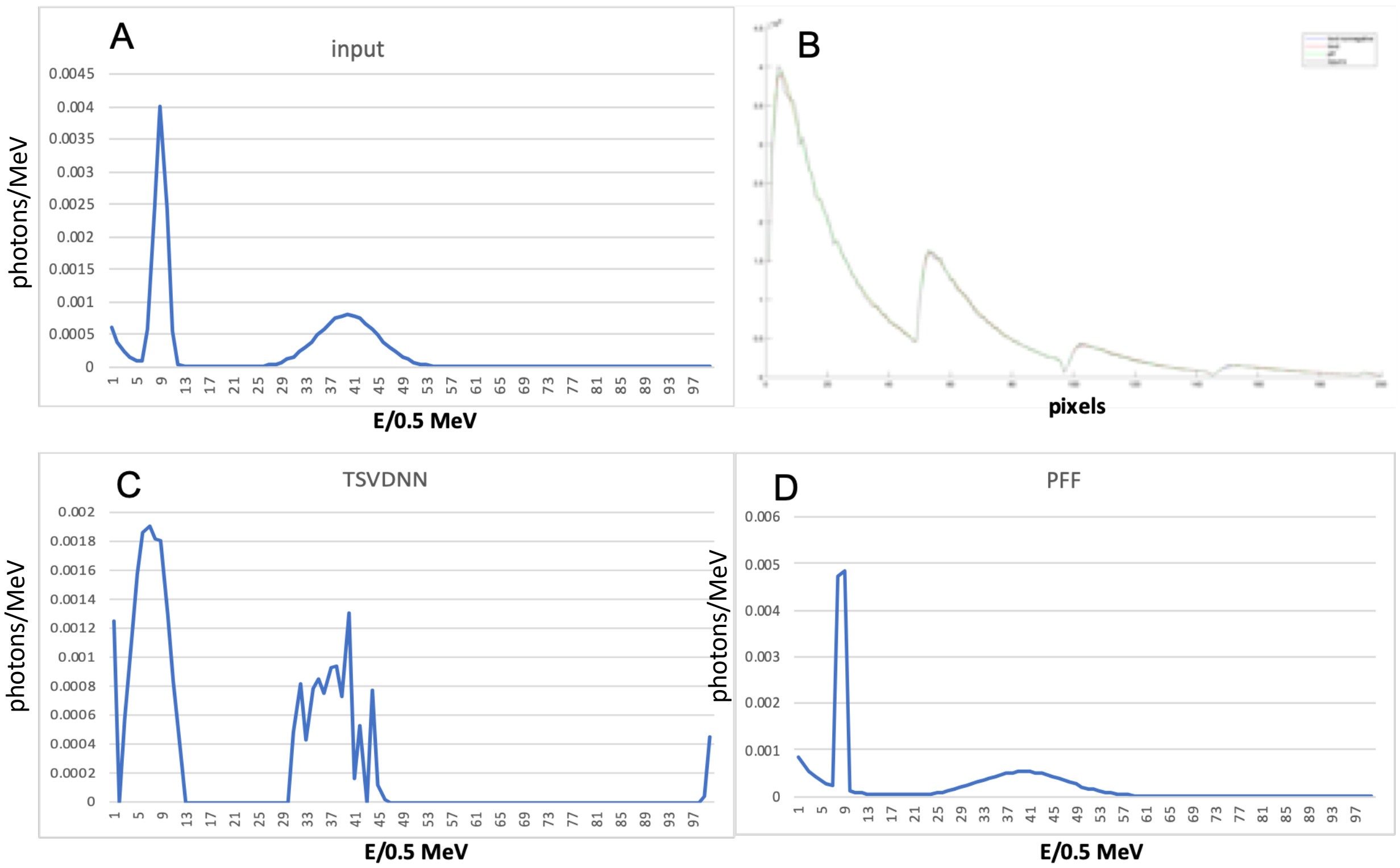

**FIG.11** (a) Model input spectrum consisting of an exponential function plus a narrow line at 4.5 MeV, and a Gaussian "bump" at 20 MeV. This model spectrum is convolved with the DRM to generate the 200-pixel light profile of the black curve in Fig.(b) after 5% random noise is added to the data. (c) TSVDNN inversion solution of the black curve of Fig.(b) without any prior knowledge of input spectrum. (d) Best-fit PFF solution using an analytic model based on the overall shape of the TSVDNN solution of Fig.(c). In Fig.(b), the light output profiles generated by the TSVDNN solution (red) and the PFF solution (green) are compared to the input generated light profile (black). The units on the y-axis are arbitrary in all figures.

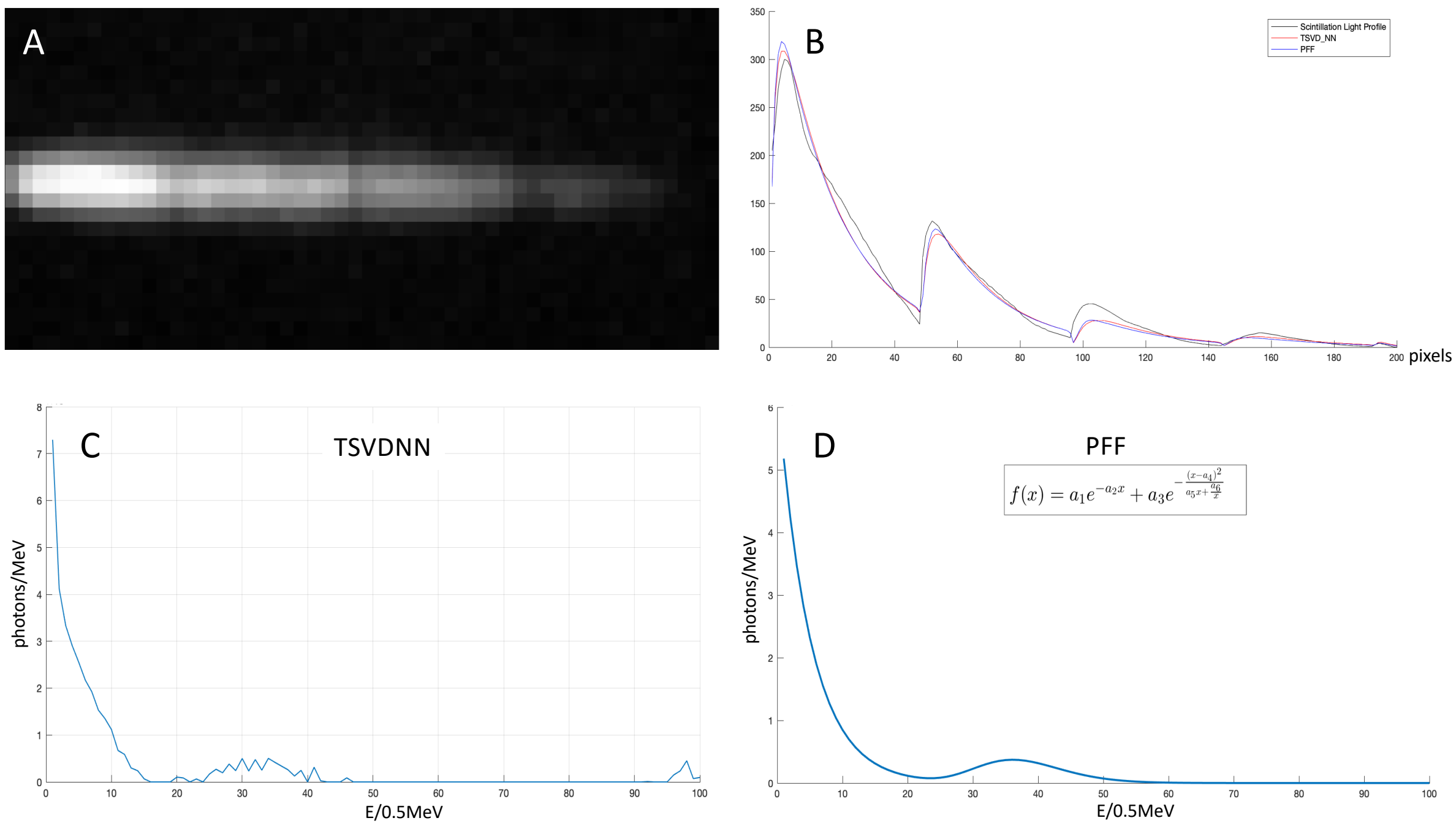

**FIG.12** (a) The SAS image of TPW Shot 10084 (Fig.3) is first reprocessed into uniform square pixels after conversion of the intensity to linear scale and background subtraction. (b) The black curve is a plot of the digitized light profile of Fig.(a) after consolidating all bright pixels into 200 "effective" pixels and correcting for defective and saturated pixels. We also plot the light profiles generated by the TSVDNN solution (red curve) and best-fit PFF solution (blue curve) to compare with the black curve. (c) A gamma-ray spectrum inverted from the black curve of Fig.(b) using the TSVDNN algorithm. (d) The best-fit PFF spectrum based on the TSVDNN spectrum of Fig.(c). Inset is the analytic model used in the PFF iteration initialized with parameters from the TSVDNN solution of Fig.(c). The y-axis units of Figs.(b)(c)(d) are arbitrary.

Fig.12 illustrates the process of extracting the incident gamma-ray spectrum from the raw SAS image of Fig.3. We first digitize the raw image into 36x48 uniform square pixels, remove the nonlinearity in the CCD camera response, and subtract the background. The reprocessed light profile (Fig.12(a)) is further improved by correcting for defective and saturated pixels to create the 200-pixel light profile data (black curve in Fig.12(b)). This light profile is first inverted using the TSVDNN algorithm without any assumption of the incident gamma-ray spectrum. The TSVDNN solution is shown in Fig.12(c). We then use an analytic model based on the TSVDNN solution as the PFF trial function (Fig.12(d) inset) and initiate the PFF iteration with parameters based on the TSVDNN solution. The final best-fit PFF solution is given in Fig.12(d). This PFF solution consists of a low-energy exponential component with decay energy kT=2.5 MeV, plus a broad bump centered at ~18 MeV. The low-energy exponential spectrum is consistent with results from our FSS spectrometer below 6 MeV, but the broad bump at ~ 18 MeV is a new spectral feature which could not have been discovered using the FSS spectrometer. We detected such 15-20 MeV gamma-ray bumps in a subset of TPW shots when the SAS lies in the forward cone between laser forward and target normal directions (see discussions in Sec.6).

Fig.13 shows two other examples of gamma-ray spectra obtained from our 2016 TPW experiments which consist of only an exponential component without any high energy bump, using the same algorithm and procedure as in Fig.12. The best-fit PFF exponential decay energies are kT=1.36 MeV for Shot 10028 and kT=1.52 MeV for Shot 10121. These kT values are consistent with those obtained from our FSS spectrometers for similar laser targets (Fig.14, kT=1.6 MeV). By comparing the raw SAS images of Fig.13 with that of Fig.3, we see that the biggest differences are the lack of emission beyond longitudinal pixel y ~ 28 and the rapid narrowing of the transverse light profile in Fig.13, while Fig.3 shows abundant emission beyond y = 28 and much broader emissions in the transverse dimension. These differences are telltale signatures of light emission by higher energy gamma-rays, consistent with the expectations based on the GEANT4 simulations shown in Fig.8 (yellow color contour).

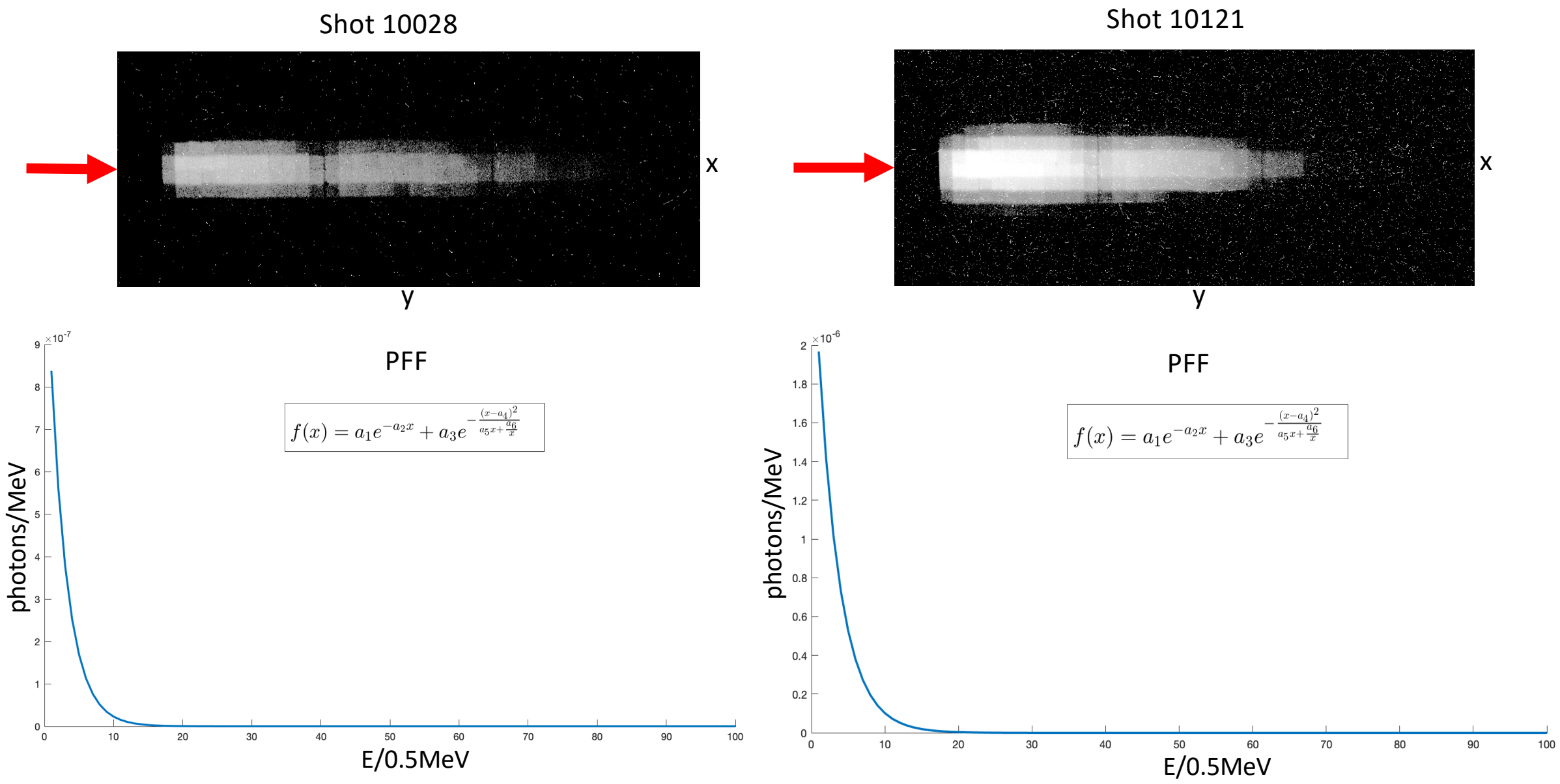

**FIG.13** Two SAS images from 2016 TPW experiments which give rise to pure exponential gamma-ray spectra using our hybrid TSVDNN+PFF algorithm. Red arrows indicate incident gamma-ray beams. The numerical algorithm used to obtain these best-fit PFF solutions are identical to those used for Fig.12. Top panels are raw SAS images and bottom panels are the best-fit PFF solutions. Insets are the PFF trial functions based on the TSVDNN solutions. The best-fit PFF values are: $a_2$= 0.37 for Shot 10028, $a_2$ = 0.33 for Shot 10121 and $a_3$ = 0 in both cases.

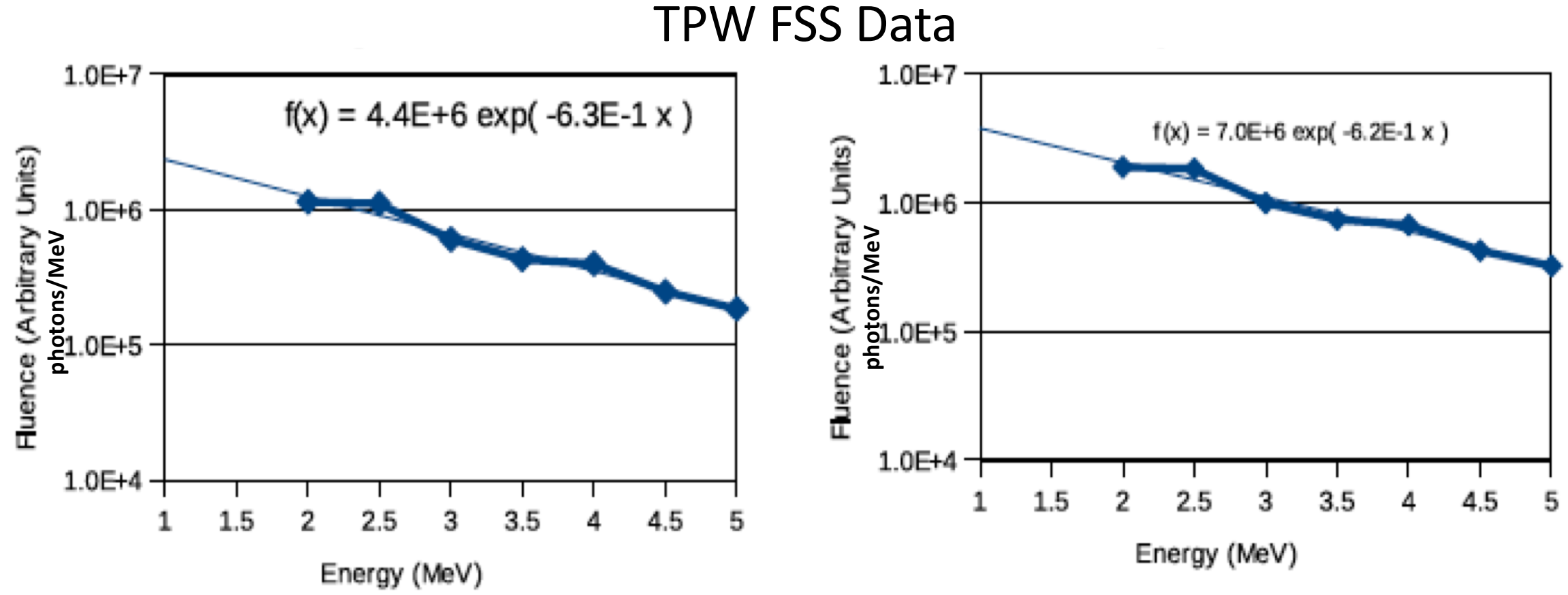

**FIG.14** Sample FSS spectra obtained in our TPW experiments for targets similar to those in

Fig.13. Their exponential decay energies (=1.6 MeV) are consistent with the SAS results of Fig.13. The FSS results help to validate the SAS results up to 5 MeV.

In general, we find that the gamma-ray spectra deconvolved from the SAS images using our hybrid TSVDNN+PFF algorithm are robust and stable, even when we increase the background noise and systematic errors. General agreement of the low-energy exponential spectra obtained from the SAS with those obtained from the FSS validates both the SAS approach and our inversion method based on the hybrid TSVDNN+PFF algorithm. Statistical and systematic errors from the digitization and reprocessing of the SAS light profile data are inevitable. We believe that the key to the stability and reliability of our inverted gamma-ray spectra from SAS images is the 2D nature plus the large number of bright pixels of the SAS light profile, which tightly constrains the inverted spectrum. This is further helped by reducing the DRM to a rectangular matrix of 200x100, so that the inversion problem becomes “over-determined”. For the first time in ultra-intense short-pulse laser experiments, the SAS allows us to accurately measure the gamma-ray spectrum far above 6 MeV with 0.5 MeV resolution. One new discovery enabled by the SAS is the existence of a gamma-ray “bump” peaking around 15-20 MeV in a fraction of our TPW shots, when the SAS lies in the forward cone between the laser forward and target normal directions [20][21].

## VI. DISCUSSIONS AND SUMMARY

Fig.12 shows that in a fraction of our TPW shots, the gamma-ray spectrum exhibits a “bump” peaking at 15-20 MeV which has not been reported before. However, since no reliable high-resolution gamma-ray spectral measurements have been made in this energy range before, it is unclear if this gamma-ray bump may be present in other short-pulse ultra-intense laser experiments. The origin and physical explanation of this bump is a subject of intense ongoing research, which lies beyond the scope of this paper. Here it suffices to state that (a) we are highly confident of the reality of this bump based on the analysis of many TPW shots, and detailed modeling of the detector response shows that it is not an artifact caused by software or hardware deficiencies, (b) the bump is absent for gamma-ray spectra measured with the SAS in directions lying outside the TPW forward cone [20][21], (c) the peak energy of the bump is consistent with the peak energy of TPW hot electrons [22]. At present, a tantalizing but unproven hypothesis under active investigation is that the bump may be produced by Compton upscattering of multi-keV x-rays under special conditions.

Besides calibration using low-energy radioactive sources and validation using the FSS, it is obviously desirable to calibrate the SAS at higher energies (> 6 MeV) using gamma-rays from electron accelerators. Our original plan to do this at nearby medical centers was interrupted by the retirement of our medical collaborators plus COVID19 shutdowns. But we hope to restart such calibration work in the near future. However, the bremsstrahlung gamma-ray spectra emitted by electron accelerators are broadband and model dependent. So the calibration is not as clean-cut as using mono-energetic gamma-rays of known energy. However, the only mono-energetic high-energy gamma-ray sources are relativistic radioactive ion beams and relativistic positron annihilation-in-flight beams, neither of which are currently available to us.

In this paper we have presented the basic concept, design, response matrix, deconvolution algorithm and sample results of a new type of gamma-ray spectrometer called the SAS, which combines the principles of attenuation and scintillation by a finely pixelated LYSO scintillator matrix. The key innovation is the use of mm-scale pixelated LYSO(Ce) crystals with fully reflective coating so that the light emerging from every LYSO pixel faithfully represents the local energy deposition within that pixel. The LYSO matrix provides high-quality 2D images of scintillation light emitted by gamma-rays up to 50 MeV in a compact volume. The abundant light output of LYSO crystals provides high signal-to-noise images with a large number of bright pixels even at low gamma-ray fluence, and the high-speed CCD camera is capable of imaging scintillation light patterns with millisecond exposures, ideal for high-repetition-rate laser applications. We have demonstrated that the SAS can produce reliable spectrum with 0.5 MeV resolution, for incident gamma-rays containing both narrow spectral lines and broad continuum features. Enlarging the LYSO matrix to incorporate more longitudinal pixels (e.g. 36x72) will extend the gamma-ray spectrum to 100 MeV. Alternatively, we can also increase the spectral resolution, especially at low energies, by using smaller pixels (e.g. 1mm x 1mm instead of 1.5mm x 1.5mm). Deconvolution of all gamma-ray spectra obtained from our past experiments is in progress. They will be published in future papers once the analysis is completed.

## ACKNOWLEDGEMENTS

This work was supported by DOE grants DE-SC0021327 and DE-SC0016505.

## AURHOR DECLARATIONS

### Conflict of Interest

The authors have no conflicts to disclose.

## DATA AVAILABILITY

High level processed data that support the findings of this paper are available from the corresponding author upon reasonable request.